\documentclass{article}

\usepackage{arxiv}
\usepackage[utf8]{inputenc} 
\usepackage[T1]{fontenc}    

\usepackage{graphicx}
\usepackage{amssymb}
\usepackage{amsmath}
\usepackage{mathrsfs}
\usepackage[square,sort,comma,numbers]{natbib}

  \let\geq=\geqslant

\providecommand\bnabla{\boldsymbol{\nabla}}
\providecommand\bcdot{\boldsymbol{\cdot}}

\newcommand\etab{\boldsymbol{\eta}}
\newcommand\etabp{\dot{\boldsymbol{\eta}}}

\newcommand\Sp{\ensuremath{S_{\parallel}}}
\newcommand\Sn{\ensuremath{S_{\perp}}}
\newcommand\eb{\ensuremath{\boldsymbol{e}}}
\newcommand\nb{\ensuremath{\boldsymbol{n}}}
\newcommand\ebp{\ensuremath{\dot{\boldsymbol{e}}}}

\newcommand\xb{\ensuremath{\boldsymbol{x}}}
\newcommand\ub{\ensuremath{\boldsymbol{u}}}
\newcommand\gradu{\ensuremath{\bnabla\ub}}
\newcommand\xib{\ensuremath{\boldsymbol{\xi}}}
\newcommand\dif{\ensuremath{\text{d}}}
\newcommand\xc{\ensuremath{x_\text{c}}}
\newcommand\yc{\ensuremath{y_\text{c}}}
\newcommand\xsb{\ensuremath{\xb_\text{s}}}

\newcommand\xs{\ensuremath{x_\text{s}}}
\newcommand\ys{\ensuremath{y_\text{s}}}

\newcommand\ts{\ensuremath{\theta_{\text{s}}}}
\newcommand\uc{\ensuremath{U_\text{c}}}
\newcommand\uw{\ensuremath{U_\text{w}}}
\newcommand\uo{\ensuremath{U_0}}
\newcommand{\xin}{\xb_0}
\newcommand\ppuu{\ensuremath{p_1^{\gamma_1}}}
\newcommand\ppud{\ensuremath{p_2^{\gamma_1}}}
\newcommand\pput{\ensuremath{p_3^{\gamma_1}}}

\newcommand\ppdu{\ensuremath{p_1^{\gamma_2}}}
\newcommand\ppdd{\ensuremath{p_2^{\gamma_2}}}
\newcommand\ppdt{\ensuremath{p_3^{\gamma_2}}}
\newcommand\ppdq{\ensuremath{p_4^{\gamma_2}}}
\newcommand\Rey{\mbox{\textit{Re}}}

\newcommand\eg{e.g.}
\newcommand\ie{i.e.}
\newcommand{\bib}[1]{}

\newcommand\pa{(\textit{a})}
\newcommand\pb{(\textit{b})}
\newcommand\pc{(\textit{c})}
\newcommand\pd{(\textit{d})}
\newcommand\pe{(\textit{e})}
\newcommand\pf{(\textit{f})}
\newcommand\pg{(\textit{g})}
\newcommand\ph{(\textit{h})}
\newcommand\pab{(\textit{a,\,b})}
\newcommand\pcd{(\textit{c,\,d})}
\newcommand\pef{(\textit{e,\,f})}

\newcommand{\prods}[3]{\ensuremath{\left\langle #1 ,(#2)#3 \right\rangle}}
\newcommand{\prodss}[3]{\ensuremath{\left\langle #1 ,#2#3 \right\rangle}}
\newcommand{\xo}{{\xb(t)}}
\newcommand{\conj}[1]{\ensuremath{{#1}^*}}

\newcommand{\fig}[1]{figure~\ref{fig#1}}
\newcommand{\figs}[1]{figures~\ref{fig#1}}
\newcommand{\Fig}[1]{Figure~\ref{fig#1}}
\newcommand{\Figs}[1]{Figures~\ref{fig#1}}

\newcommand \LADYF{\normalsize L\kern -.35em\lower -.5ex\hbox{\scriptsize A} \normalsize\kern -.4em\lower .2ex
	\hbox{\footnotesize D}\kern -.25em\lower .ex\hbox{Y}\kern -.1em 
	\hbox{\footnotesize F}}

\title{Toward the detection of moving separation in unsteady flows}

\author{
  Philippe Miron\thanks{Email address for correspondence: philippe.miron@polymtl.ca}\\
  Department of Mechanical Engineering\\
  \LADYF, Polytechnique Montr{\'e}al,\\Montr{\'e}al QC\\
  H3C 3A7, Canada\\
   \And
 J{\'e}r{\^o}me V{\'e}tel\\
  Department of Mechanical Engineering\\
  \LADYF, Polytechnique Montr{\'e}al,\\Montr{\'e}al QC\\
  H3C 3A7, Canada\\
}

\begin{document}
\maketitle

\begin{abstract} 
In many engineering systems operating with a working fluid, the best efficiency is reached close to a condition of flow separation, which makes its prediction very crucial in industry. Providing that wall-based quantities can be measured, we know today how to obtain good predictions for two and three-dimensional steady and periodic flows. In these flows, the separation is defined on a fixed line attached to a material surface. The last case to elucidate is the one where this line is no longer attached to the wall but on the contrary is contained within the flow. This moving separation is probably, however, the most common case of separation in natural flows and industrial applications. Since this case has received less attention during the past few years, we propose in this study to examine some properties of moving separation in two-dimensional, unsteady flows where the separation does not leave a signature on the wall. Since in this framework separation can be extracted by using a Lagrangian frame where the separation profile can be viewed as a hyperbolic unstable manifold, we propose a method to extract the separation point defined by the Lagrangian saddle point that belongs to this unstable manifold. In practice, the separation point and profile are initially extracted by detecting the most attracting Lagrangian coherent structure near the wall, and can then be advected in time for following instants. It is found that saddle points, which initially act as separation points in the viscous wall flow region, remarkably preserve their hyperbolicity even if they are ejected from the wall toward the inviscid region. Two test cases are studied, the creeping flow of a rotating and translating cylinder close to a wall, and the unsteady separation in the boundary layer generated by a planar jet impinging onto a plane wall.
\end{abstract}


\section{Introduction}

Boundary layer separation remains one of the most important unsolved, at least partially, problem in fluid mechanics, and probably one of the most important to solve considering its critical effects on many engineering systems. Since the pioneering work of Prandtl in 1904 on two-dimensional steady separation, a considerable number of ideas have emerged to capture unsteady separation in two and three dimensional flows.

A large number of studies have initially focused on solving boundary layer equations, thus defining the separation as the state when the solution to these equations become singular \citep{Sears1975}. This same postulate was also used by \cite{VanDommelen1990} to capture the unsteady boundary-layer separation but in Lagrangian coordinates, a representation that better reveals the nature of separation and offers more advantages than in the Eulerian frame. However, the occurring of a singularity in the numerical integration of boundary layer equations means that the approximation upon which they are based become invalid, and this is not necessarily related to separation. The reader is referred to the recent reviews of \cite{Ruban2011} and \cite{Cassel2014} for further details.

A consensus on a general criterion to detect unsteady separation was not found until the seminal work of \cite{Haller2004} who provided a kinematic theory for two-dimensional flows. In this entirely new, non-linear and Lagrangian approach, separation appears as a material instability induced by an unstable manifold (defined in finite or infinite time) that emanates from the wall at a boundary point. In forward time, the unstable manifold attracts and ejects particles from the wall, and the theory provides the location and shape of the separation profile. One of the most remarkable results is that this approach can be applied to flows with general time dependence, as well as in compressible flows, thus unifying prior works into a single theory and extending them to cases not covered before.

\cite{Haller2004} defined two types of separation. A \textit{fixed} separation occurs when the flow has a well-defined mean value, such as in periodic flows where the separation characteristics are easy to obtain since integration in time is applied over one single period. In this case, the boundary point of separation is not moving on the wall but is fixed at a location where the backward-time average of the skin friction vanishes (weighted by a function of the fluid density if compressibility effects are present). The time-dependence then appears only in the shape of the separation profile, mainly in the angle of separation. These results have since been extended to three-dimensional steady flows \citep{Surana2006}, then to three-dimensional unsteady flows \citep{Surana2008a}.

The treatment of \textit{moving} separation is more delicate. This case occurs when the separation point may move, or may appear and disappear in the flow. Since classical invariant manifold theory cannot apply, \cite{Haller2004} used finite-time unstable manifolds \citep{Haller2000} to capture the moving point, thus yielding to non-unique moving separation profiles. In a further contribution, \cite{Surana2008b} proposed a criterion to capture separation in slow-fast systems, where the mean and fluctuating flow components are characterised by different time scales, such as in the wake of a circular cylinder placed in a time-varying crossflow. In both works, the moving separation point is defined at the location where the time-varying mean skin friction vanishes, and requires first to extract the temporal mean flow components, which is done numerically, for example by using wavelet-based denoising methods.

Despite the important potential impact of these recent works, a very few studies have focused on their experimental validation. We believe this is mainly due to two reasons. The first one comes from difficulties to measure required quantities. In the case of fixed separation, while the time-independent location of the separation on the wall is determined from time-averaged on-wall measurements of shear stress, pressure and their spatial derivatives, the time-dependent separation profile (slope) inside the flow requires the knowledge of instantaneous values of these quantities. The wall pressure signature is not so easy to obtain accurately if high frequencies are present but this is feasible with carefully performed experimental procedures. On the contrary, the skin friction is still challenging to measure accurately, all the more so a high spatial resolution is also needed to obtain the derivatives along the wall. To the authors knowledge, only one experimental study has been reported in the literature. \cite{Weldon2008} studied a rotating cylinder whose axis can be oscillated parallel to a wall, thus manipulating an unsteady separation. Under periodic, quasi-periodic and random forcing, observations reveal that separation emanates from a fixed location on the surface, its position and orientation over time being accurately predicted by the theory. However, this unique example concerns a slow viscous flow in the quasi-steady Stokes regime (based on the cylinder diameter and the circumferential velocity, the Reynolds number $\Rey$ is less than 1). Moreover, numerical simulations were used to provide missing information that could not be obtained from experiments, which only provided flow visualisations.

As a consequence, separation criteria have mainly been validated only with computational flow models. A large majority of studies used different variants of an unsteady separation bubble model derived by \cite{Ghosh1998} or other flow models derived from Taylor series expansion solutions of the Navier--Stokes equation provided by the algorithm of \cite{Perry1986} \citep{Haller2004, Kilic2005, Surana2006}. In a sense, since in these examples the entire flow is computed from data at the wall, it seems reasonable to think that this should also be the case for the separation that must leave an imprint on the wall. More sophisticated flow models, obtained from full numerical solutions of the Navier--Stokes equation, are used in \cite{Surana2007} and \cite{Surana2008a}, but none of them reflects flows dominated by vortex shedding and, as a consequence, turbulence. As a modest contribution, we have adapted the formulae of \cite{Haller2004} in cylindrical coordinates to predict separation and attachment profiles in the vicinity of a circular cylinder \citep{Miron2015a}, confirming that the alternated K\'arm\'an vortex street falls into the category of fixed separation, the profiles of which are captured by the theory.

While separation can be easily captured from numerical simulations data in laminar flows, it turns out to be very challenging with turbulence: the flow behaviour in the vicinity of boundaries should not be influenced by wall modelling, as it is the case in large eddy simulation, and for direct numerical simulation, the computation of separation characteristics requires long calculation times and therefore significant computational resources to obtain statistical convergence of mean flow quantities. This explains the second reason why so few studies have aimed at examining the validity of the kinematic theory of separation through experiments: to date, most of flow visualisations seem to indicate that with turbulence, flow separation is always moving. We have however to consider that the region where the separation profile is attached to the wall at a fixed point is so indistinguishably small that an experimental detection is difficult to obtain \citep{Weldon2008}. Some scepticism is also present when experimental measurements of the vanishing wall shear is found to be moving, but this is related to the fact that in the aerodynamics community, this criterion is still often considered as the separation location, whereas it is known from a long time that vanishing wall shear `does not denote separation in any meaningful sense in unsteady flow' \citep{Sears1975}.

More importantly, it is probable that moving separation is more common than fixed separation in unsteady flows, this latter phenomenon appearing exclusively in periodic and quasi-periodic flows. In turbulent flows for instance, the time scale of the smallest eddies is much lower than that of the coherent structures, which in turn is much smaller than that of the mean flow, meaning that the separation point is probably not fixed spatially. However, a very few studies address the detection of moving separation points. The theoretical background used in \cite{Haller2004} and \cite{Surana2008b} has two limitations. First, the flow is supposed to admit fast temporal fluctuations superimposed on a slower time-varying, \ie\ a mean, flow that is extracted numerically (for example through a low-order polynomial least-squares fit or a wavelet-based decomposition), but this time scale separation is not clear in some flows, or does not exist at all. In turbulence for instance, the spectral content of velocity fluctuations is continuous. Second, the moving separation is defined at the location where the wall shear of the mean flow vanishes, but as we shall see, this is not always true.

The objective of this paper is to contribute to a better understanding of unsteady moving separation. In particular, by defining the separation point as a point off the wall acting as a Lagrangian saddle point that moves close to the wall, the unstable manifold to which it belongs to corresponds to the separation profile. To detect separation, this profile is first captured by detecting a hyperbolic Lagrangian coherent structure (LCS) in backward time. In two dimensions, this LCS represents  a material line that exhibits locally the strongest attraction from all nearby material lines \citep{Haller2011}. The position of the separation point can then be inferred from the measure of the Lagrangian rate of strain \citep[see, \eg,][]{Haller2002} along the LCS. In a large variety of applications, the instantaneous rate of strain along LCSs is examined to analyse different flow phenomena, such as the detection of hyperbolic cores in two-dimensional turbulence \citep{Mathur2007}, the birth of secondary vortices from hairpin vortices in turbulent boundary layers \citep{Green2007}, or the prediction of imminent shape changes in oil spills in the sea \citep{Olascoaga2012}. In \cite{Lekien2008}, the analysis of the rate of strain is also used to detect separation on a slip boundary that replaces the LCS. They show that the instantaneous value is not sufficient to build the detection criterion, especially when random flow fluctuations and noise are present. Fortunately, as the boundary is everywhere tangent to the local velocity vector, fluid trajectories in its vicinity can be explicitly solved and integrated in time from on-wall flow quantities. This leads to the computation of the backward-time average of the rate of strain that better reveals the location of the separation point as well as the angle of the separation profile. We show here that the reason why trajectories can be computed is that the boundary acts as a natural material barrier, and we extend this tool to any material line evolving in the interior of the flow to detect separation that occurs off the wall.

The paper is organised as follows. General observations on fixed and moving separation phenomena are first described in \S\,\ref{des}. Theoretical developments based on previous studies are then detailed in \S\,\ref{the}. This is followed by the presentation of the results in \S\,\ref{res}, namely the wall flow near a rotating and translating cylinder, and the two-dimensional, planar impinging jet, before drawing the conclusion in \S\,\ref{con}.

\section{Examples of unsteady separation induced by a rotating cylinder}\label{des}

The base flow that is used in the first part of this article to illustrate the separation phenomena is the two-dimensional velocity field of the creeping flow developing around a rotating circular cylinder moving close to a plane wall \citep{Klono2001}. The solution, detailed in Appendix~\ref{appA}, was chosen over the separation bubble model derived by \cite{Ghosh1998} because it is valid not only in the vicinity of the wall but also everywhere inside the flow, a requirement here since we shall see that the separation point is captured above the wall. Moreover, this is a simple configuration where separation can be completely controlled.

\begin{figure}
  \centerline{\includegraphics{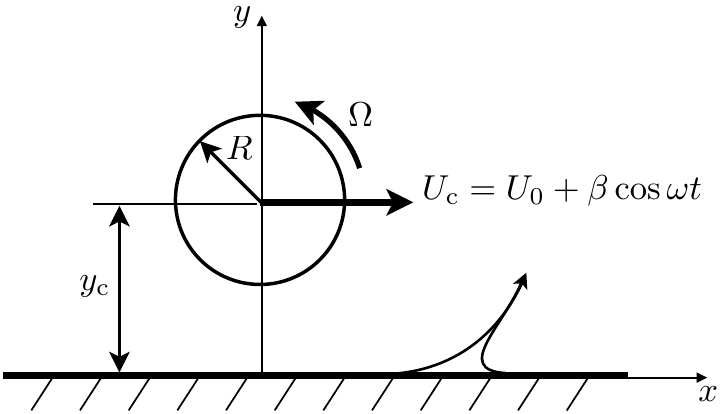}}
  \caption{Flow geometry and parameters of the rotating and translating cylinder test case.}
  \label{fig01}
\end{figure}

The flow geometry is presented in \fig{01}. A cylinder of radius $R$, initially at position $\xc=0$ and $\yc$, rotates about its axis at a constant angular velocity $\Omega$, which leads to the appearance of a separation point at a position on the wall downstream of the cylinder. By translating the cylinder on a line parallel to the wall following a prescribed trajectory, the separation can be manipulated under different flow conditions. Throughout the article, we use $R=1$, $\yc=2$ and $\Omega = 1$, and the cylinder velocity is $\uc = \uo + \beta\cos \omega t$, where $\uo$ is a constant translating velocity over which an oscillating movement of angular frequency $\omega$ and amplitude $\beta$ is superimposed. Before examining the moving separation, the fixed separation is first investigated.

\subsection{Fixed separation}

If the cylinder oscillates without translation ($\uo =0$), the flow is periodic of period $T$ and falls into the category of fixed separation. The separation occurs at a point on the wall where the skin friction, averaged over one time period, vanishes, and the temporal separation profile can be computed using formulas given in \cite{Haller2004}.

\begin{figure}
  \centerline{\includegraphics{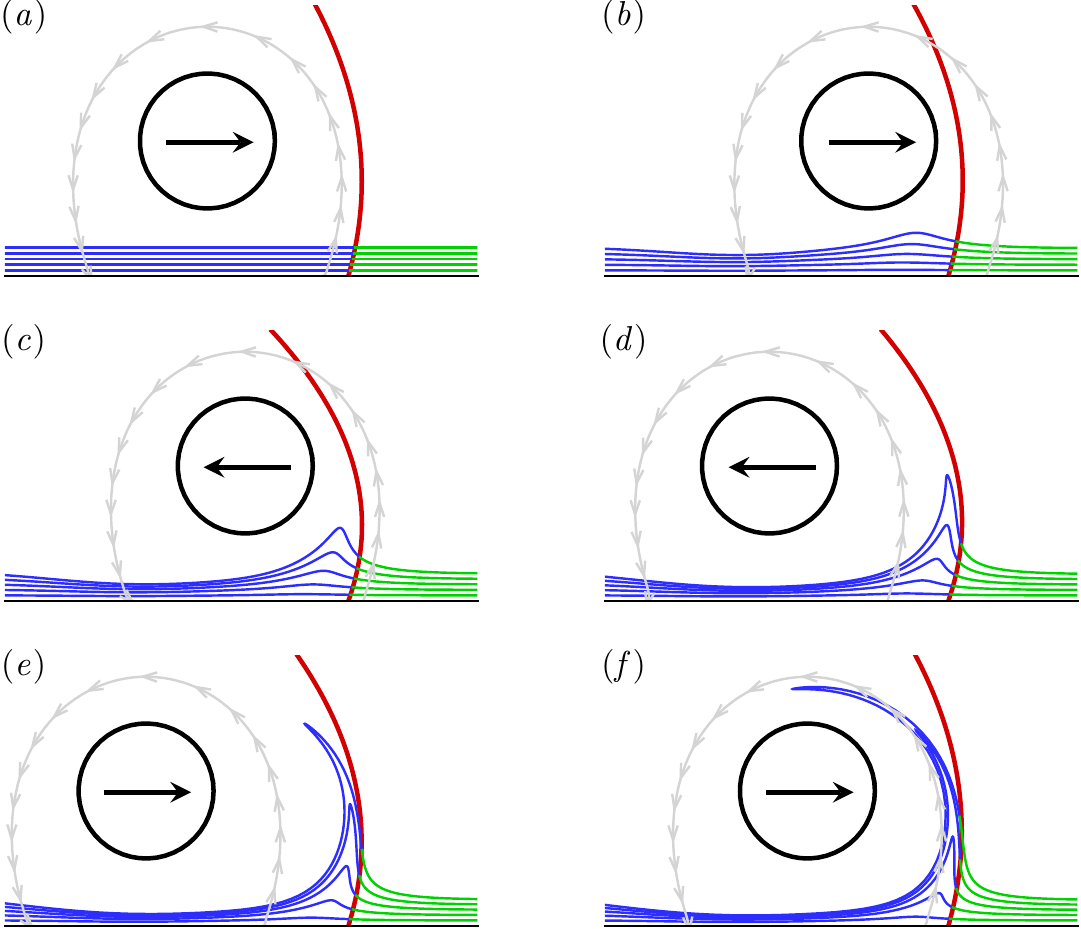}}
  \caption{Fixed unsteady separation for the time-periodic flow with $\uo  =0$, $\omega = 2\pi$, and $\beta =6$: \pa\ $t=0$, \pb\ $t=3.2\,T$, \pc\ $t=6.4\,T$, \pd\ $t=9.6\,T$, \pe\ $t=12.8\,T$ and \pf\ $t=16\,T$. The Lagrangian separation profile (red) is shown with the instantaneous streamline (grey) emanating from the zero skin friction points. Fluid particles are initially placed on material lines aligned with the wall using two different colours (blue and green) depending on their location with regard to the predicted separation profile at the initial time $t=0$.}
  \label{fig02}
\end{figure}

\Fig{02} presents the results obtained with $T=1$ and $\beta =6$. The second-order separation profile was obtained over time and compared to the instantaneous streamline emanating from the instantaneous zero skin friction point. We can note that fluid particles, initially aligned with the wall, separate from it as they are advected in time to form a spike that is located upstream of the actual separation point that indeed appears to be fixed. This contrasts with the position of the instantaneous zero skin friction point that oscillates around the true separation point, which illustrates that the separation mechanism is not related to the Prandtl criterion. We may further note that the time-dependent separation profile prediction agrees with the particle motion close to the wall. As the distance from the wall increases, the prediction degrades but we must note that the theory provides the separation profile in terms of an approximation obtained from a series expansion in $y$. Only a second-order separation profile has been computed here, and as a consequence a better agreement would have been obtained by including higher order terms.

\subsection{Moving separation}\label{mov}

To generate a moving separation, the cylinder can be translated. As a first illustrative example, the velocity of the cylinder is set to a constant value ($\uo  =0.3, \beta = 0$) with results presented in \fig{03}. Fluid particles are released from two lines parallel to the wall at the initial time $t=0$ (\fig{03}a). Their positions are then followed in time in a reference frame moving with the cylinder. As for the fixed separation, they gradually separate from the wall to form a spike whose location does not coincide with the instantaneous zero skin friction point. However, while particles located on the blue lines are ejected away from the wall, particles that are initially located at a lower transverse location (green lines) form a tip but stay close to the wall from which they are never ejected. This is the case even at large time instants as for example at $t=100$ in \fig{03}\pf.

\begin{figure}
  \centerline{\includegraphics{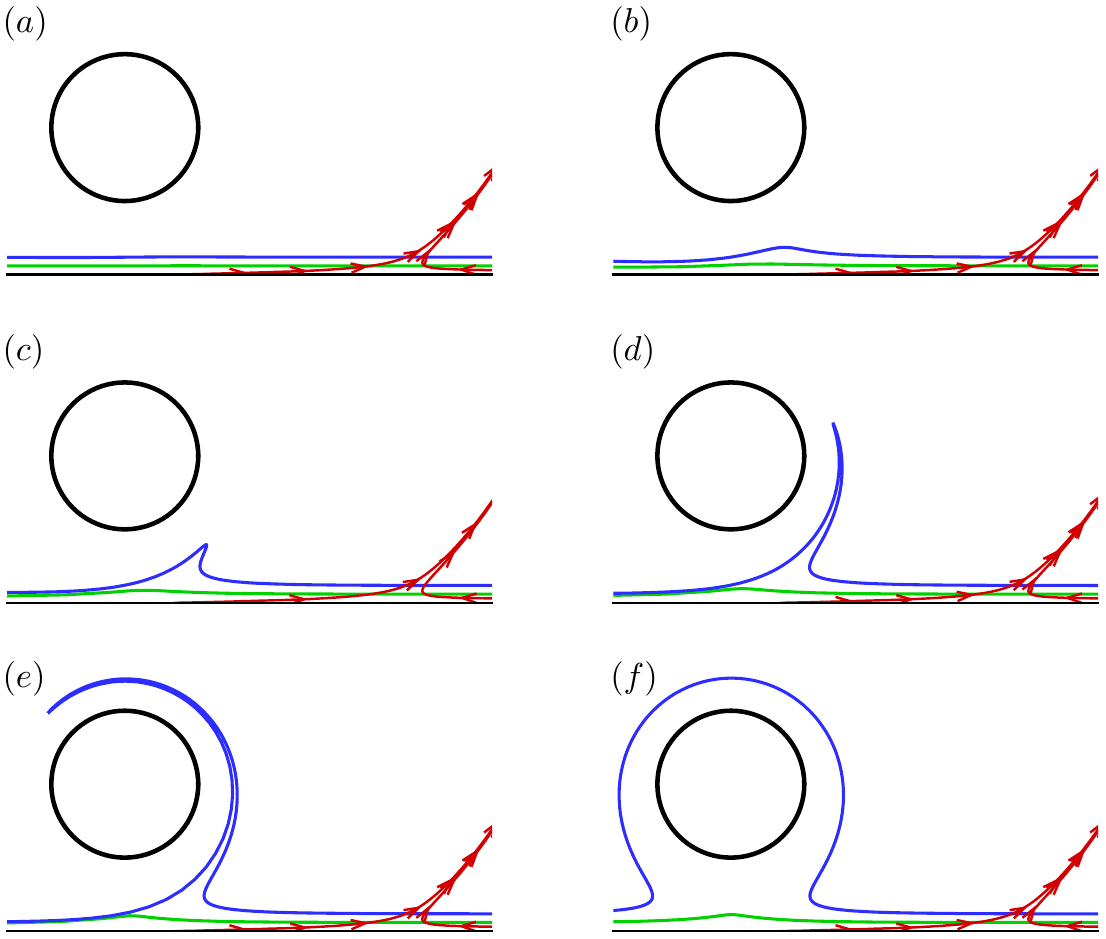}}
  \caption{Moving separation with $\uo  =0.3$ and $\beta = 0$. Material lines (blue and green) initially aligned with the wall at $t=0$ in \pa\ are then advected in time and seen in a reference frame moving with the cylinder at times \pb\ $t=4$, \pc\ $t=8$, \pd\ $t=12$, \pe\ $t=16$ and \pf\ $t=100$. Instantaneous streamlines are shown in red (note that in \pf, the blue line show material points that have been attracted and ejected from the wall by separation, then wrapped around the cylinder and finally transported to its left side; since other points are left behind the cylinder, they end up being disconnected from the material line).}
  \label{fig03}
\end{figure}

If we follow \cite{Surana2008b}, there is no flow fluctuation in this example, and hence separation is predicted to appear where the wall shear of the mean flow, i.e. of the instantaneous flow here, vanishes, which is not observed as no separation occurs near the zero skin-friction point. Moreover, it appears difficult to find a criterion to detect separation using only quantities measured along the wall but on the contrary results seem to go, at least qualitatively, in the direction of the Moore--Rott--Sears (MRS) principle \citep{Moore1958, Rott1956, Sears1956} stating that unsteady separation takes place at a point off the boundary. This latter criterion further states that separation occurs at a moving point where the wall component of the shear is zero, i.e. inside the boundary layer rather that on the wall, thus requiring to know the velocity of the separation point. However, in the example above, the wall shear does not vanish anywhere in the vicinity of the spike formation, indicating that quantitatively, the MRS principle is not verified. However, the MRS principle can only be applied in the context of boundary-layer theory, \ie\ in the asymptotic limit when the Reynolds number $\Rey \rightarrow \infty$, which is not the case here. Therefore, another detection criterion is required.

\section{Detection of separation}\label{the}

Instead of seeking the separation point on a boundary, it appears unavoidable to look for a point inside the flow, i.e. off the boundary. From observations of results shown in \fig{03}, and in accordance with the MRS principle, the separation should be more adequately described by a saddle point located above the wall (as sketched in \fig{04} and detailed below). This is, for example, the case for the flow around a moving cylinder \citep[see, \eg,][]{Koromilas1980}.

\subsection{Theoretical background}\label{set}

The theoretical background used in this study is based on the work of \cite{Haller2003} and \cite{Lekien2008}. As presented in Appendix~\ref{appB}, these works used a local change of coordinate, whose axis are the tangent and normal unit vectors $\eb(\xb(t),t)$ and $\nb(\xb(t),t)$ to the local velocity vector, to study Lagrangian hyperbolicity. For example, if we take at an arbitrary initial time a material line that is everywhere tangent to $\eb$, we obtain a streamline. When advected in time, this line is no longer everywhere tangent to $\eb$ except in particular cases. In steady flows, streamlines coincide with trajectories, so that the initial streamline is, in subsequent times, always tangent to $\eb$. As shown in Appendix~\ref{appB}, an interesting consequence resulting from this property is that an infinitesimal perturbation to the trajectory of a point belonging to the streamline can be explicitly computed in time (through the term $\alpha$ in system~\ref{sys} that vanishes), thus allowing the study of the stability of the streamline. As a final result, the $\lambda$-criterion, that characterises the cumulative rate of the strain on the line instead of the instantaneous value of this quantity, can be used to characterize different phenomena. In \cite{Haller2003}, it serves to detect high stretching regions in barotropic turbulence more efficiently that for example the finite-time Lyapunov exponent. In the case of a boundary with slip velocity conditions, the wall is always tangent to the velocity vectors, and therefore the $\lambda$-criterion can be used to detect separation on free-slip walls, such as current separation on coasts from surface ocean velocity data for applications in geophysics \citep{Lekien2008}.

\begin{figure}
  \centerline{\includegraphics{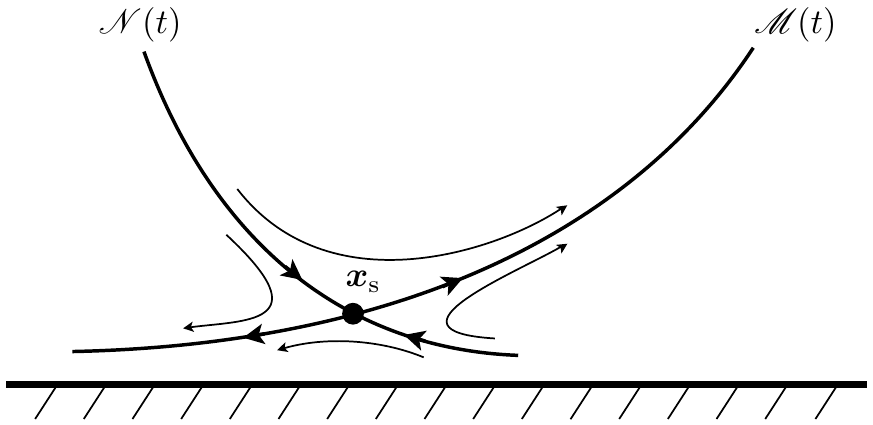}}
  \caption{Separation pattern defined around a Lagrangian saddle-point of coordinates $\xsb =(\xs,\ys)$, including an attracting material line (unstable manifold) $\mathscr{M}(t)$ and a repelling material line (stable manifold) $\mathscr{N}(t)$.}
  \label{fig04}
\end{figure}

Here, the problem is reversed. The material line is not imposed but on the contrary is what we are looking for. \Fig{04} presents schematically the trajectories of fluid particles around a Lagrangian saddle point moving close to a wall. The separation point, located at $\xsb$, is the intersecting point between an attracting material line $\mathscr{M}(t)$, that coincides with the separation profile, and a repelling material line $\mathscr{N}(t)$, corresponding to an unstable and a stable manifold, respectively. Since $\mathscr{M}(t)$ is a material line, by defining $\eb(t_0)$ (respectively $\nb(t_0)$) as unit vectors along and tangent (respectively normal) to $\mathscr{M}(t_0)$ (\ie\ not necessarily aligned with the local velocity vectors) at an arbitrary initial time $t_0$, then $\mathscr{M}(t)$ remains everywhere tangent to the family of unit vectors $\eb(t)$ by advection in time. To simplify, if at $t_0$ a small element of this line is $\xib(t_0) = \xi(t_0)\eb(t_0)$, then $\xib(t)$ is always on this line for $t\geq t_0$, i.e.:
\begin{align}\label{eq:cond}
\xib\bcdot\nb = 0 \qquad \text{and} \qquad \frac{\dif}{\dif t}(\xib\bcdot\nb) = 0,
\end{align}
which gives, using the linearised flow \eqref{lin},
\begin{align}\label{eq:alpha}
\prods{\nb}{\gradu}{\eb}_\xo - \nb\bcdot\ebp |_\xo = 0.
\end{align}
Since \eqref{eq:alpha} simply means $\alpha(t) = 0$, we are in the situation where the $\lambda$-criterion can be used to detect separation. By setting $\eb =(\cos\theta, \sin\theta)$ and $\nb =(-\sin\theta, \cos\theta)$, \eqref{eq:alpha} also provides an equation for $\theta(t)$ depending on the spatial derivatives of velocity components:
\begin{align}\label{eq:theta}
\dot{\theta} = v_x \cos^2 \theta  - u_y \sin^2 \theta + (v_y-u_x)\cos\theta\sin\theta .
\end{align}
For example, in the case of a separation point fixed on a wall, $\mathscr{N}(t)$ would be the wall boundary ($y=0$), and in the case of an incompressible flow, at the wall we would have $u_x=-v_y=0$ and $v_x = 0$. This would lead to
\begin{align}
\frac{\dif}{\dif t} \left(\frac{1}{\tan\theta}\right) = u_y,
\end{align}
which is the result obtained by \cite{Haller2004}.

In the case of the cylinder translating at a constant velocity \uo\ described in \S\,\ref{mov}, since the flow is steady in the frame moving with the cylinder, the separation point and profile should also be steady in this reference frame. As a consequence, the separation point has a constant streamwise velocity \uo\ and zero transverse velocity. Since the streamlines have a reflectional symmetry about the vertical axis passing through the centre of the cylinder, on this axis we have $v=0$ which indicates a probable location of the separation point ($\xs=\xc$). Equation~\eqref{eq:flow} was then solved with $u=\uo$ and $v=0$ to find the coordinate $\ys$ of the separation point. As for the separation line, $\dot{\theta} = 0$, and since at the separation point we have $u_x=v_y=0$ due again to the reflectional symmetry of the streamlines, \eqref{eq:theta} gives
\begin{align}\label{eq:angle}
\tan\theta = \pm\left(\frac{v_x}{u_y}\right)^{1/2}.
\end{align}
In the reference frame moving with the cylinder the flow is steady, the separation point has zero relative velocity components, and then also corresponds to a critical point according to the concepts of \cite{Perry1987}. As a consequence, the slope angle given by \eqref{eq:angle} corresponds to the direction of the eigenvectors of the velocity gradient tensor.

\begin{figure}
  \centerline{\includegraphics{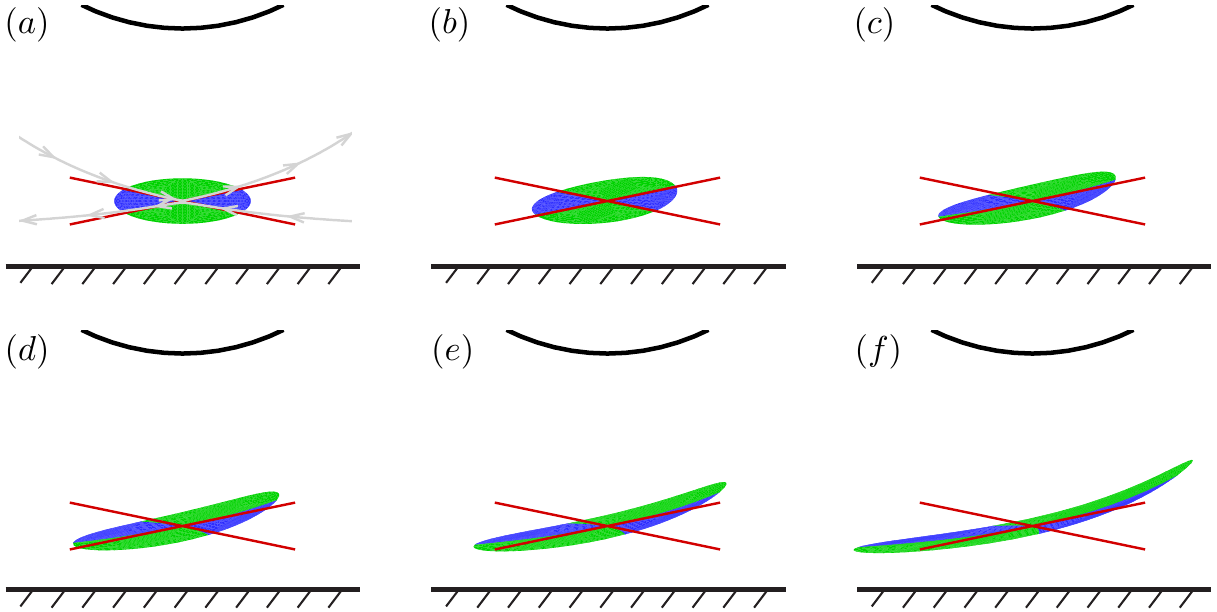}}
  \caption{Prediction of moving separation with $\uo  =0.3$ and $\beta = 0$: \pa\ $t=0$, \pb\ $t=0.8$, \pc\ $t=1.6$, \pd\ $t=2.4$, \pe\ $t=3.2$ and \pf\ $t=4$. Particles (seen in a reference frame moving with the cylinder) are released with a colour (blue or green) depending on their initial location relative to the red straight lines that correspond to the tangent to the stable and unstable manifolds predicted by equation~\eqref{eq:angle}. Grey lines are instantaneous streamlines (computed in the moving frame) passing close to the separation point.}
  \label{fig05}
\end{figure}

\Fig{05} presents the separation mechanism by advecting fluid particles initially located in an elliptical region centred on the predicted separation point. In the close vicinity of the saddle-point, particles trajectories remain tangent to the stable and unstable manifolds as they are convected by the mean flow. Particles in the upper quadrant are ejected from the wall while in the lower quadrant they are attracted towards the wall. This example is however quite classical since the streamline pattern, coincident with particle trajectories since the flow is steady in the frame of reference linked to the cylinder, is the standard pattern induced by a convected vortex close to a wall \citep{Doligalski1994}. If the flow displays an arbitrary time-dependence, streamlines have no relationship to Lagrangian material lines that therefore need to be detected. This is examined in the next subsection.
 
\subsection{Initial conditions}\label{ic}

Before capturing the separation point, the unstable manifold at time $t_0$, $\mathscr{M}(t_0)$, is first extracted by detecting hyperbolic Lagrangian coherent structures (LCSs) in the vicinity of the wall. Hyperbolic LCSs were traditionally defined geometrically as local maximizing curves (defined as ridges) of the finite-time Lyapunov exponent (FTLE) field with particular properties, but it was however shown since that such definitions are inadequate \citep[see, \eg, the recent review of][for details]{Haller2015}. Recently, \cite{Haller2011} refined the definition of LCSs and proposed a more consistent and convenient way to extract them based on their physical properties. With this new approach, a hyperbolic LCS is locally the strongest repelling or attracting material line over a finite time interval \citep[note that a geodesic theory, generalizing the concept of LCSs to hyperbolic, elliptic and parabolic material lines, is presented in][but in the remainder of the paper, the term LCS will refer to a hyperbolic LCS as other types will not be used]{Haller2012}. More precisely, the LCS is captured by maximizing a finite-time normal repulsion measure over all nearby material lines. Therefore, if we adopt this definition, $\mathscr{M}(t_0)$ can be seen as a hyperbolic material line computed in backward time.
The numerical algorithm followed here to extract LCSs is fully detailed in \cite{Farazmand2012}. Briefly, we compute \textit{strainlines} that, by definition, are material lines that are everywhere tangent to the field of unit eigenvectors associated with the smaller eigenvalue field of the Cauchy--Green strain tensor. In practice, strainlines are simply computed as trajectories of the eigenvector field for a given time, in a way similar to the computation of streamlines from the velocity field. The procedure starts with the selection of initial coordinates that verify two criteria. The first criterion ensures that the normal repulsion rate computed along the LCS \citep[defined in][]{Haller2011} is larger than the tangential stretching rate, and the second that it reaches locally a maximum relative to other material lines. Strainlines are then integrated until either the boundary of the domain is reached, or if one of the criteria fails during the integration. From all initial grid points respecting these conditions, a family of Lagrangian coherent structures is first obtained, typically of a few hundreds in complex flows and high mesh density. In fact, many of those structures are very close to each others and only differ because of their different initial position and the numerical errors introduced during the integration. To filter out similar LCSs and facilitate the analysis, the average value of the FTLE along all candidates are compared against each others in a close circular region (typically define by a small radius). Those with the highest values are selected and others discarded. In our case, this method effectively decreases the number of LCSs of more than one order of magnitude \citep[for further technical details, see][]{Farazmand2012}. Once a material line is extracted, its evolution can be followed in time by advecting fluid particles that compose it, performed in the present study by using a 5th order explicit Runge--Kutta integration scheme. This computation can be implemented without difficulty because the forward time advection of an attracting LCS is numerically stable.

Knowing the geometry of the LCS, the separation point can be obtained by seeking the hyperbolic (saddle) point that belongs to $\mathscr{M}(t_0)$, \ie\ the point with the highest tangential rate of strain (or lowest normal rate of strain). In practice, the $\lambda$ exponent, defined by \eqref{lambda}, is computed for all points of the LCS over the time interval $[t_0,t]$ and $\xsb (t_0)$ is obtained where $\lambda$ shows a maximum that can be detected whenever the integration time is long enough. Note that since at $t_0$ the LCS is defined by the desired number of points, an adequate precision on the initial unit vectors can be obtained to compute the initial value of $\lambda$. However, when the LCS is advected in time, this precision can decrease as particles separate from each other. This difficulty is avoided by using \eqref{eq:theta} instead to compute the exact angle in time, since $\theta(t)$ can be simply obtained from the knowledge of its initial value and of the velocity spatial derivatives. This contrasts with previous methods used in the literature \citep[see, \eg,][]{Farazmand2012, Olascoaga2012} that require to compute the gradient of the flowmap at each time instant. Finally, once the maximum of $\lambda$ is extracted at $t$, the location of the separation point $\xsb(t_0)$ can be traced backward in time, as well as the initial separation angle $\ts(t_0)$.

\section{Results}\label{res}

The detection of the separation point and angle shown in the previous section will be illustrated on two different flows. The general procedures will be detailed in Sec.~\ref{res1} with a rotating cylinder that can translate and oscillate in a direction parallel to a plane wall, and will then be applied to a vortical flow in Sec.~\ref{res2}.

\subsection{The rotating and translating cylinder}\label{res1}

\begin{figure}
  \centerline{\includegraphics{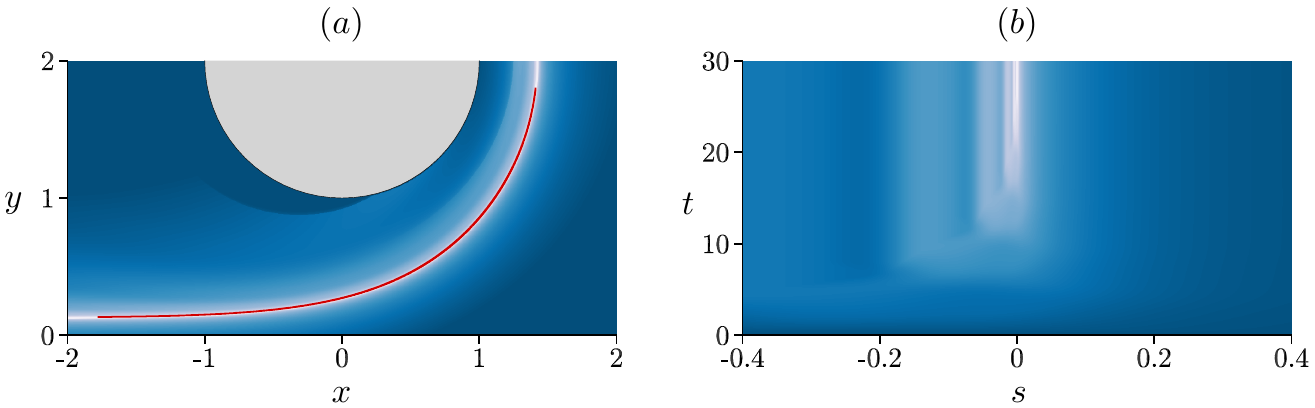}}
  \caption{Extraction of the separation point and angle for the flow with $\uo  =0.3$, $\beta = 0.5$ and $\omega = 2\pi/5$. In \pa, contours of the FTLE field are visualised (levels increase from blue to white) together with the LCS (red) extracted at $t=0$ with the algorithm presented in \cite{Farazmand2012}. $\lambda = \int_{0}^t \Sp(\tau)\mbox{d}\tau$, computed as a function of $s$, the initial curvilinear coordinate on the LCS at $t=0$ normalised by the length of the LCS, is visualised in \pb\ through isocontours (from blue to white). The origin of $s$ corresponds to the point on the LCS in \pa\ where $x=0$.}
  \label{fig06}
\end{figure}

The first example considered here corresponds to the Stokes flow introduced in Sec.~\ref{mov}, defined by a translating and rotating cylinder, but with, in addition, a periodic perturbation with $\beta = 0.5$ and $\omega = 2\pi/5$. The extraction of the separation point and angle at $t=0$ is presented in \fig{06}. The FTLE field, computed in backward time, is shown in \fig{06}\pa\ and clearly highlights the existence of a LCS emanating from the wall that wraps around the cylinder. The Cauchy--Green strain tensor computed to obtain the FTLE field was then used to extract the exact geometry of the LCS obtained from the algorithm presented in \cite{Farazmand2012}. In this simple flow, only one LCS was extracted. This is a particular case since generally a large number of hyperbolic lines is obtained (see \S~\ref{res2}). As can be seen, the line is indeed localised close to the extrema of the FTLE field, with a spatial resolution as high as desired. \Fig{06}\pb\ shows the computation of $\lambda$ along the LCS when advected in time as a function of the initial curvilinear coordinate $s$ and the integration time $t$. With increasing time, a maximum clearly appears that can be traced back to $t=0$ to find the initial location of the saddle point on the LCS that corresponds to the separation point. This point is found at $s=0$, \ie\ at $x=0$ on the LCS identified in \fig{06}\pa. This confirms that the separation point is clearly located above the wall.
 
\begin{figure}
  \centerline{\includegraphics{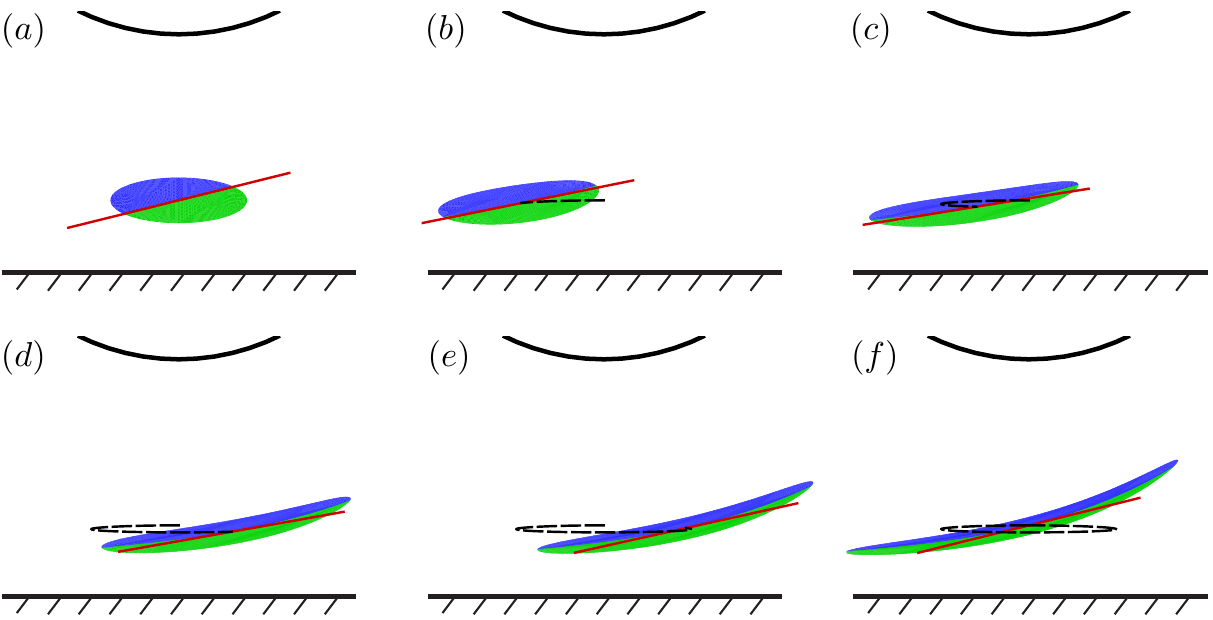}}
\caption{Prediction of moving separation with $\uo  =0.3$, $\beta = 0.5$ and $\omega = 2\pi/5$: \pa\ $t=0$, \pb\ $t=T/5$, \pc\ $t=2T/5$, \pd\ $t=3T/5$, \pe\ $t=4T/5$ and \pf\ $t=T$. The legend is the same as in \fig{05}. The dotted line indicates the trajectory of the separation point.}
  \label{fig07}
\end{figure}

Once the initial location and angle of the separation point are known, its evolution in time can be predicted. Results are presented in \fig{07}. As observed, fluid particles globally follow the trajectory of the separation point without crossing the predicted separation profile in the close vicinity of the separation point (this is not the case in the periphery because  the profile is approximated by the tangent defined at the separation point). One can note that the trajectory of the separation point is disconnected from that of the cylinder, \ie\ the separation point does not move with the cylinder but oscillates close to it following an ellipsoidal path. On the contrary, the Eulerian saddle point, as identified in \fig{05}\pa, is always located on the vertical axis passing through the cylinder centre due again to the reflectional symmetry of streamlines about the vertical axis passing through the cylinder centre. This illustrates that in the current example an Eulerian-based approach cannot be used to capture unsteady separation that is Lagrangian by nature.

\begin{figure}
  \centerline{\includegraphics{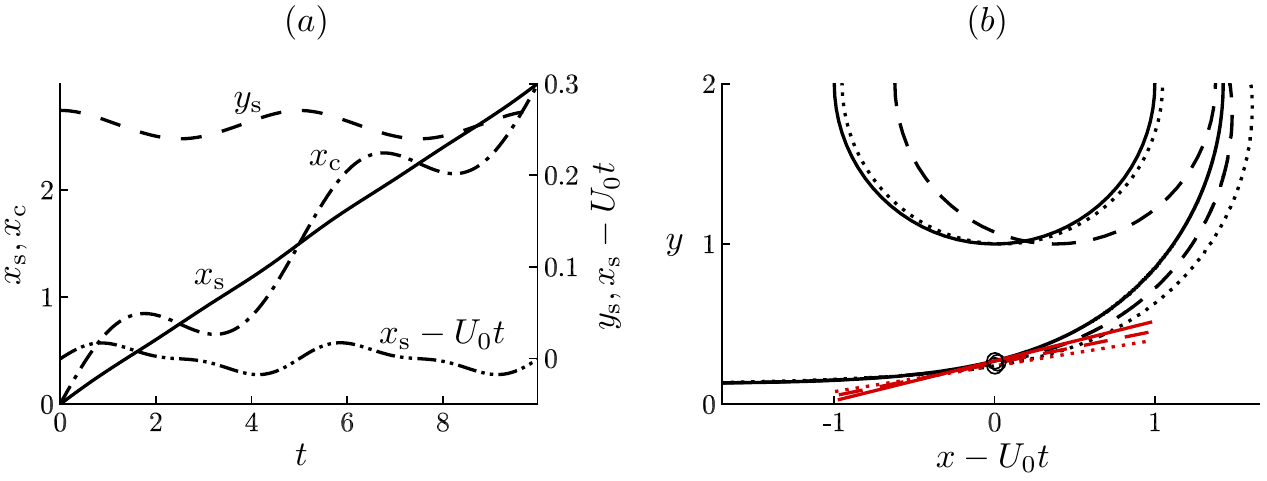}}
  \caption{Tracking of the separation point and of the separation profile over time. In \pa, the coordinates of the separation point are plotted in different ways, and in particular are compared to that of the cylinder. \pb\ shows the position of the LCS and of the cylinder for three time instants in a reference frame moving at the velocity $\uo$. The separation point is in indicated by the circles and the theoretical prediction of the slope of the separation profile at the separation point is indicated in red: --------, $t=0$ ; $---$, $t=1$; $\cdot\cdot\cdot\cdot\cdot$, $t=2.4$.}
  \label{fig08}
\end{figure}

Further details on the separation point kinematics is provided in \fig{08}. In \fig{08}\pa, it can be observed that the position $\xs$ differs from the linear motion $\uo t$ by only a slight value. In \fig{08}\pb, the separation point and angle are visualised in a frame moving at a velocity $\uo$ for three time instants. One can note that the angle of separation changes over time but only slightly. These results are consistent with those obtained for the fixed separation in the sense that the Lagrangian saddle points exhibit less oscillations than their Eulerian counterpart.

\subsection{The impinging jet}\label{res2}

The flow detailed in the previous section is a quite simple Stokes flow. To increase the level of complexity, we study here the vortical flow generated by the impingement of a laminar jet on a plane wall perpendicular to the main jet velocity. This configuration has been the subject of many studies due to its fundamental and industrial importance, especially in cooling or heating systems. The most active research topic probably concerns the local maxima of the Nusselt number observed at two different radial locations when the plate is heated, the origin of which has been discussed in numerous works \citep[see, \eg][for a recent literature review]{Dairay2015}.

\begin{figure}
  \centerline{\includegraphics[scale=0.8]{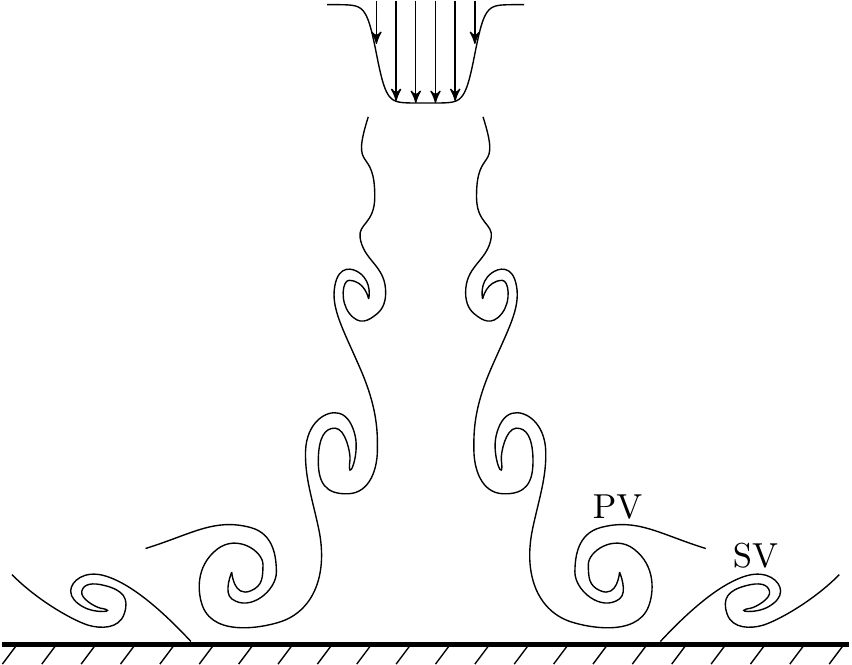}}
  \caption{Physical mechanism of the unsteady separation in an impinging jet with PV and SV denoting the primary and the secondary vortex, respectively \citep[adapted from][]{Didden1985}.}
  \label{fig09}
\end{figure}

The geometry, while simple, exhibits a lot of complex mechanisms. One of them concerns the formation of an unsteady  separation phenomenon. As schematically presented in \fig{09}, when a primary vortex (PV) created in the mixing layer of the jet approaches the wall, an adverse pressure gradient is induced close to the wall, yielding to the formation of a shear layer. This shear layer is unstable, and leads to the generation of a secondary vortex (SV), counter-rotating with the primary vortex \citep[a thorough understanding of this phenomenon is detailed in][]{Dairay2015}. Simultaneously, an unsteady separation is generated in the near region of the wall, which then moves downstream with the primary vortex. This scenario, while periodic, involves a separation that appears and disappears, and therefore falls into the category of \textit{moving} separation according to \cite{Haller2004}. In the experiments conducted by \cite{Didden1985} on a forced jet, it is found that the wall shear stress is at any time positive in the vicinity of the secondary vortex formation, thus invalidating the use of the zero skin friction point to detect separation.

In the present study, post-processing tools presented in previous sections are applied to the case of a two-dimensional, planar impinging jet. Flow conditions are simple but sufficient to be dynamically relevant of an unsteady separation. The flow arrangement is similar to the one shown in \fig{09}, \ie\ the jet is vertical and oriented toward the bottom. Finite element method was used to compute the velocity fields. The distance between the jet nozzle and the plane wall is 10\,$D$, where $D$ is the jet width. The computational domain is $50D$ in the axial direction parallel to the plate, symmetrical to the jet centreline, and $10D$ in the transverse direction. A no-slip boundary condition is imposed on the wall and on the upper boundary. At the exit sections, an outflow condition that minimise the influence of the domain truncation is imposed \citep{dong2014}. To trigger the jet instability, a hyperbolic tangent velocity profile (shown in \fig{09}) is imposed at the jet nozzle:
\begin{equation}
  U(x)=-\frac{U_\text{a}}{2}\left[1+\tanh \left( 5-\frac{10|x|}{D} \right) \right]
\end{equation}
where $x=0$ is located on the symmetry axis of the jet and $U_\text{a}$ is the velocity on the jet axis. The mesh, composed of Q2-Q1 elements, is refined as we approach the wall. The Reynolds number is defined by $\Rey = U_\text{a}D/\nu$, where $\nu$ is the kinematic viscosity, and is fixed at $\Rey = 500$.

\begin{figure}
  \centerline{\includegraphics{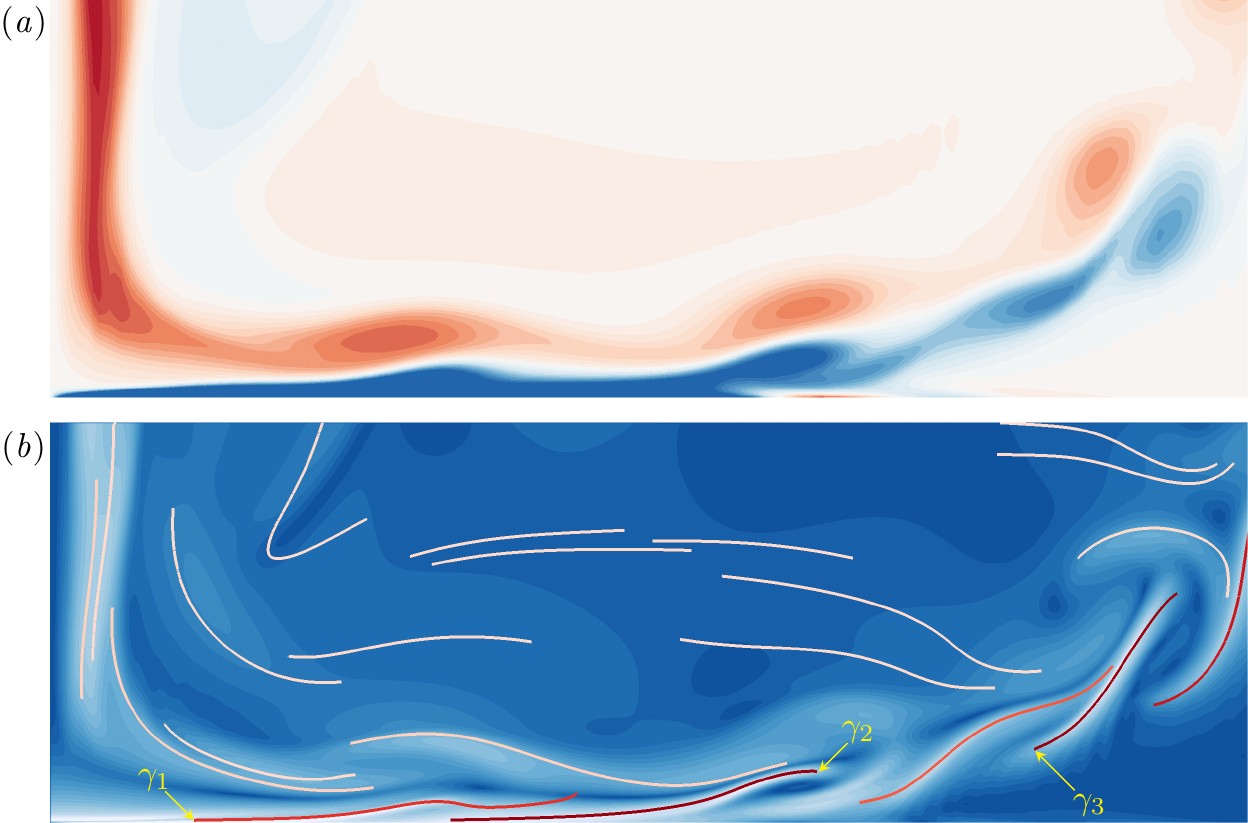}}
  \caption{Visualisation of the region impinged by the vertical jet. \pa\ Contours of the out-of-plane vorticity (the blue and red colours correspond to a clockwise and a counter-clockwise rotation, respectively). \pb\ LCS material lines coloured by the FTLE averaged along each line (lower to higher values increase from light to dark red) superimposed on the FTLE contours (from blue to white).}
  \label{fig10}
\end{figure}

\Fig{10} shows a view of the flow region where secondary vortices are generated. In \fig{10}\pa, the primary vortices are highlighted by negative vorticity contours (counter-clockwise rotation), while their passage close to the wall generates a local shear layer of vorticity with an opposite sign (clockwise rotation). When the shear layer instability is sufficiently developed, secondary vortices appear downstream and lift up from the wall, generating an unsteady separation.

\Fig{10}\pb\ shows contours of the FTLE field together with extracted LCSs using the algorithm detailed in \cite{Farazmand2012}. We can note that the LCSs are not necessarily ridges of the FTLE field, and conversely, local maxima of the FTLE do not necessarily correspond to a LCS \citep{Haller2011}. However, as indicated by contours of the average FTLE along LCSs, most attracting LCSs are close to local maxima of the FTLE. These LCSs are dynamically linked to the formation and development of vortices seen in \fig{10}\pa. In particular, two of them, denoted as $\gamma_1$ and $\gamma_2$, are detected close to the wall, and contain a section that is adjacent to the wall.

\begin{figure}
  \centerline{\includegraphics{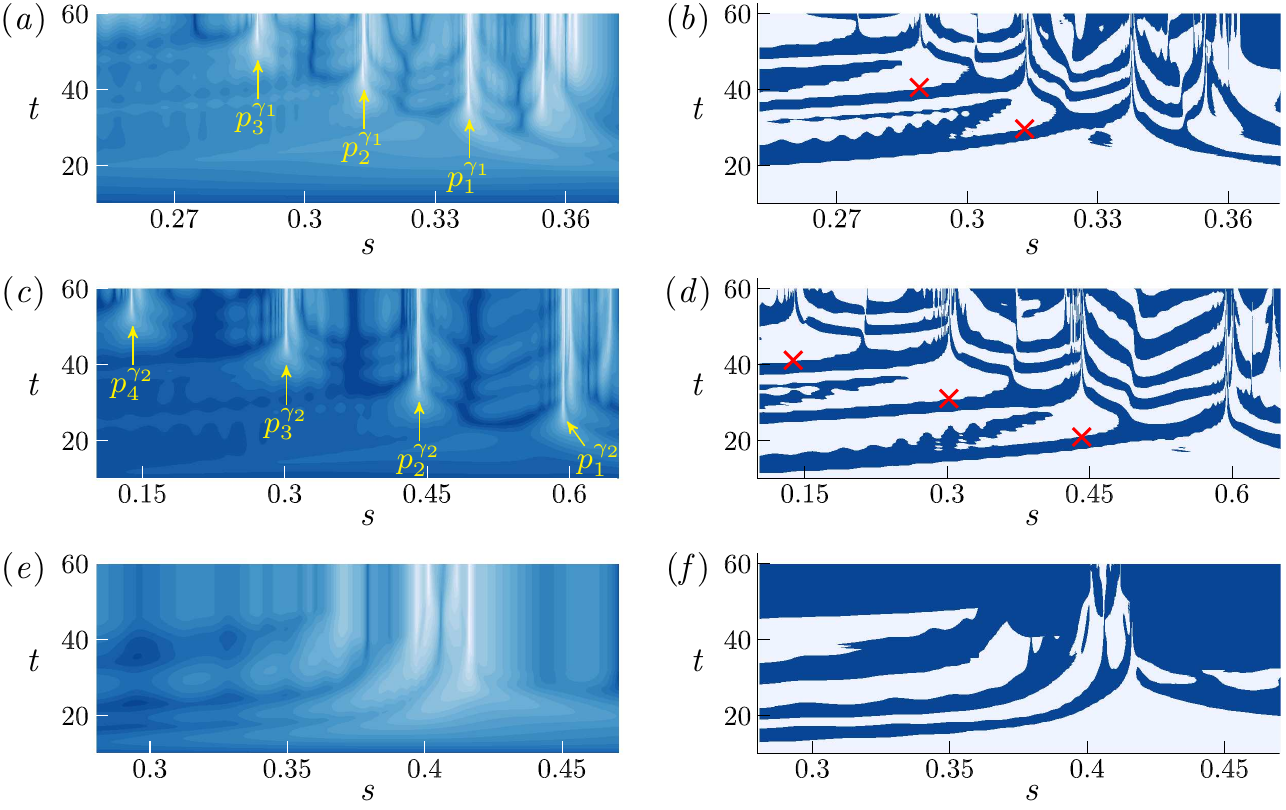}}
  \caption{Contours of $\lambda = \int_{0}^t \Sp(\tau)\mbox{d}\tau$ (left plots) and instantaneous values of the tangential strain rate $\Sp$ (right plots) as a function of the normalised curvilinear coordinate $s$ (same definition as in the legend of \fig{06}, oriented from left to right) and the integration time $t$. \pab\ correspond to the LCS identified by $\gamma_1$ in \fig{10}\pb, \pcd\ to $\gamma_2$ and \pef\ to $\gamma_3$. Red crosses indicate the instants from which $\Sp$ at the points $p_i^{\gamma_j}$ preserves a positive sign.}
  \label{fig11}
\end{figure}

The analysis of LCSs highlighted in \fig{10} can be further deepened in \fig{11}\pa\ that presents the evolution of $\lambda$ computed on $\gamma_1$ as a function of the integration time. As $t$ increases, a first point on $\gamma_1$, noted $\ppuu$, is detected. If $t$ is further increased, a second point ($\ppud$), located upstream of $\ppuu$, emerges, and then a third point ($\pput$), located upstream of the two first ones. As longer integration times are required to extract characteristic points, \figs{11}\pb\ shows a plot similar as in \fig{11}\pa\ except that the instantaneous value of the tangential strain rate is visualised instead of its time integral. Contrary to $\ppuu$ which shows a positive strain rate throughout its trajectory, $\Sp$ changes sign twice for $\ppud$ during an initial period of time before being detected. This period is even  longer for $\pput$, for which $\Sp$ experiences many sign changes. This indicates that $\lambda$ has not a monotonic variation in time and thus explains why a long integration time is necessary for its detection.

\Figs{11}\pcd\ show the same results as in \figs{11}\pab\ but for the second LCS $\gamma_2$ (visualised in \fig{10}\textit{b}). The same trends can be observed, with the detection of four saddle points on the material line ($\ppdu$ to $\ppdq$). As for $\gamma_1$, the first point is characterised by a positive rate of strain all along its history while the others experiences several sign changes. A remarkable feature of these points is that the hyperbolicity persists a long time, and even if they are ejected from the wall. As an other example, $\lambda$ of $\gamma_3$, a LCS detected far from the wall (see \fig{10}\textit{b}), was computed in the same manner as for the other LCSs. Results, shown in \figs{11}\pef, also reveal that the saddle point keeps its characteristics even when advected with the vortices in the interior of the flow.

\begin{figure}
  \centerline{\includegraphics{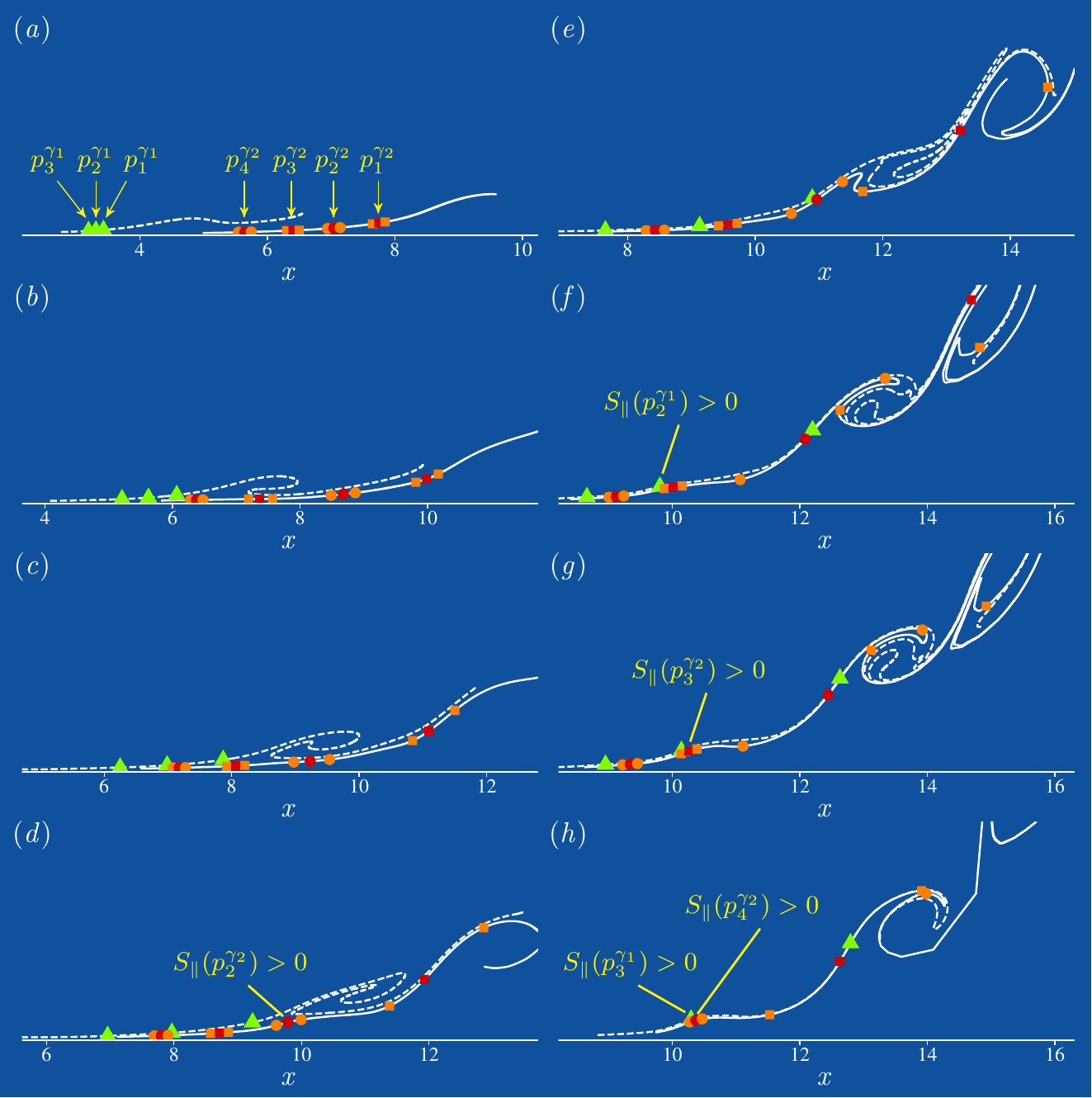}}
  \caption{Advection in time of LCSs $\gamma_1$ (dotted line) and $\gamma_2$ (solid line) detected in \fig{10}. The characteristic points extracted in \fig{11} are simultaneously followed in time at $t=10.05$ \pa, $t=14.95$ \pb, $t=18.25$ \pc, $t=21.05$ \pd, $t=25.45$ \pe, $t=29.65$ \pf, $t=30.95$ \pg\ and $t=40.65$ \ph.
  	In all plots, the three points of $\gamma_1$ ($\ppuu$ to $\pput$) are marked in green while the four points of $\gamma_2$ ($\ppdu$ to $\ppdq$) are marked in red. Orange points are points close to the characteristic points of $\gamma_2$ that are used to indicate if locally we have a stretching or a compression. Time instants where $\Sp$ becomes positive corresponds to red crosses in \fig{11}.}
  \label{fig12}
\end{figure}

An overview of the full separation mechanism is finally presented in \fig{12}. The advection of LCSs $\gamma_1$ and $\gamma_2$ is visualised in time together with the characteristic points extracted in \fig{11}. \Fig{12}\pa\ corresponds to the initial time shown in \fig{10}. In \fig{12}\pb, we can note that $\gamma_2$ is clearly stretched around points $\ppdu$ to $\ppdt$. This is still the case between \figs{12}\pb\ and \pc\ except for point $\ppdt$ that has experienced a sign inversion of the tangential strain rate since its two neighbouring points moved closer to it instead of moving away from it as between \fig{12}\pa\ and \fig{12}\pb, and as it was more formally indicated in \fig{11}\pd. Between \figs{12}\pc\ and \pd, the neighbouring points of \ppdu\ still move away from it, which confirms that $\Sp$ is always positive in \fig{11}\pd, but the point \ppdd\ has experienced a sign inversion. \Fig{12}\pd\ corresponds to the time instant for which the strain rate of \ppdd\ becomes positive and preserves its sign, as indicated by a red cross in \fig{11}\pd. This is the condition that we use to define a separation point. For subsequent times (\fig{12}e-g), we can indeed note that the separation point lifts up from the wall and is advected with the vortices, visualised by the roll up of the $\gamma_2$ curve. The characteristic points \ppuu\ to \pput\ extracted on $\gamma_1$ can also be tracked in time in \fig{12}. While $\Sp$ is positive for the furthest downstream point (\ppuu) during the entire time sequence shown, it is observed that the tangential strain rate of the second point $\ppud$ experiences several sign changes before keeping a positive value from $t=29.65$ (\fig{12}\textit{f}, and see also \fig{11}\textit{b}), which defines the time at which the separation can be defined. We can then verify that this position is very close to that of \ppdd\ in \fig{12}\pd, \ie\ of the previous separation detected on the second LCS $\gamma_2$. A  brief instant after the lift up of \ppud\ from the wall seen in \fig{12}\pf, we observe that of \ppdt\ in \fig{12}\pg, again at a close location. Finally, \fig{12}\ph\ shows data at approximately one vortex shedding period after the instant shown in \fig{12}\pg. We can observe that the most upstream saddle point of each LCS lifts up from the wall at the same instant as they are very close from each other (in reality there is a delay but it is so small that both separations are indicated on the same subfigure). Again, the locations of \pput\ and \ppdq\ before being ejected from the wall are very close to that of \ppdt\ seen in \fig{12}\pg.
 
These results show that the separation in the boundary layer developing on the wall occurs on saddle points located on LCSs moving above the wall. The bifurcation of these points from negative to positive strain rates are directly associated to the lift up of secondary vortices, which confirms the scenario proposed by \cite{Didden1985} to describe the unsteady separation at the wall. A direct consequence of this process is that a thin region exists close to the wall where the fluid is not entrained within the flow but on the contrary can even be directed towards the wall, at least for a finite-time. This reflects results obtained in \cite{ElHassan2012} where cross-correlations between wall shear stress, measured with the polarographic method, and vorticity, obtained from particle image velocimetry, was marked by a change of behaviour between the main flow and a very thin fluid layer in the near wall region.

\section{Conclusion}\label{con}

In this article, we construct a detection tool of moving separation based on post-processing of velocity fields. Separation points are defined as saddle points off the wall and can be detected by analysing an exponent (similar to a finite-Lyapunov exponent) that cumulates the history of the strain rate along their unstable manifolds. This has been used before to detect separation on slip boundaries to which the criterion was applied, but we show that the tool can also be applied to any material line within the flow. In particular, we use recent developments on LCS detection to select a family of material lines with highest normal repulsion rate \citep{Haller2011}. It is found that separation points are indeed located on these material lines and can be subsequently followed in time by the advected flow.

An analytical Stokes flow where a separation is generated by a cylinder rotating about its axis and moving parallel to a wall is first examined. If the cylinder has a pure oscillating motion, the flow is periodic and can be predicted with the exact theory of unsteady separation \citep{Haller2004}. In the case where the cylinder translates with a non-periodic motion, it is found that separation does not leave an imprint on the wall so that wall-based quantities cannot be used to reveal the phenomenon. Instead, the separation point coincides with a Lagrangian saddle point that can be captured on a hyperbolic Lagrangian coherent structure located along the wall but inside the flow. This was the general idea of the MRS principle, however this latter criterion cannot detect Lagrangian saddle points since it is based on Eulerian quantities. On the other hand, the MRS principle can be applied only to infinite Reynolds number flows, which means that the criterion is not invalidated in this study. The location and angle of separation, defined by the tangent to this LCS, can be predicted providing that they are known at an initial time, and the recipe can be generalised to any arbitrary motion of the cylinder.

To validate the methods for more complex flows exhibiting vortex development, the unsteady moving separation appearing in the vicinity of a wall impinged by a two-dimensional planar jet is investigated. Results obtained from the rotating cylinder are confirmed in the sense that separation can indeed be localised on saddle points that cyclically appear on most attracting material lines located above the wall. These separation points are then ejected in the interior of the flow and travel with vortices that also lift up from the wall.

This study is not a new theory but rather gathers findings from several theoretical works in order to deepen the understanding of moving separation. Here we can predict separation at an arbitrary time providing that we know the separation point and profile at an initial time. A self-consistent theory would be required to fully predict the unsteady separation disregarding these initial conditions, or even to fill the gap between finite Reynolds number flows and the asymptotic theory, for example in the context of the MRS principle. This is probably not unrealisable, but contrary to the case of the separation fixed on a wall, the difficulty here lies in the additional capture of the unknown separation geometry. Two other important aspects have to be considered to obtain a reliable prediction tool. The first one concerns the capture of the separation that should be based on past flow history, and the second on quantities that can be measured at the wall. Here, saddle points of material lines are captured in forward time. This could have been done in backward time but in this way the advection of an attracting LCS is numerically unstable, which requires to implement additional specific numerical methods. Concerning wall-based data, we can note that in the case of the impinging jet, Lagrangian saddle points are very close to the wall. In flows with higher Reynolds numbers, this region is very thin so that a linearised flow, based on wall quantities, could be envisaged.

\section*{Acknowledgements}
This work was supported by the Natural Sciences and Engineering Research Council of Canada (NSERC).

\appendix

\section*{Appendix A}\label{appA}

\cite{Klono2001} provided the solution to the problem of a creeping flow around a fixed rotating circular cylinder close to an infinite plane wall moving at a constant velocity. A similar flow was solved by \cite{Hackborn1997} where the cylinder is confined between two parallel infinite walls. While an analytical solution is provided, one term of the stream function is based on an integral with an infinite upper limit, which requires a numerical procedure that can be time-consuming, this is why the solution of \cite{Klono2001} was preferred here.

The original solution considered the wall at a strictly positive ordinate. This solution has been simplified here and extrapolated by setting the wall at $y=0$. If $u$ and $v$ are the velocity components in the axial and transverse directions, then the solution is given by the following complex function (with $i=\sqrt{-1}$):
\begin{align}\label{eq:flow}
u(z)-iv(z) =
&-\frac{\uw}{2\log(a)} \left[2\log \left(\frac{|{\zeta}|}{a}\right)
+ \frac{\mu}{2\zeta}(\conj{z} - z)(\zeta -i )^2 \right] \nonumber \\
& + \sigma (\zeta -i )^2 \left[  \frac{i\mu\conj{z}}{2} \left( \frac{a}{\zeta^2}
+ \frac{1}{a} \right) -  \frac{i}{\zeta}\left(a+\frac{1}{a}\right)
+\frac{1}{2a}\left(\frac{a^2}{\zeta^2}-1 \right) \right] \nonumber \\
&+ \sigma \left[a+\frac{1}{a}+i\left(\frac{a}{\conj{\zeta}} - \frac{\conj{\zeta}}{a}\right)\right],
\end{align}
where
\begin{equation}
z=x+iy, \qquad \qquad \zeta = \zeta(z)=\frac{1 + i\mu z}{i+\mu z},
\end{equation}
with the star denoting the complex conjugate.
The constants $a$, $\mu$ and $\sigma$ in \eqref{eq:flow} are obtained from the geometry and the kinematics of the cylinder:
\begin{align*}\label{eq:const}
a=\frac{R+\yc-\sqrt{\yc^2-R^2}}{R+\yc+\sqrt{\yc^2-R^2}},
\end{align*}
\begin{align*}
\mu=\frac{1}{\sqrt{\yc^2-R^2}},
\end{align*}
\begin{align*}
\sigma =\frac{a}{a^2-1}\left(
-\frac{\uw}{2\log a} + \frac{2\Omega a^2}{\mu(a^2-1)^2} \right),
\end{align*}
where $\uw$ is the velocity of the wall, $R$ is the radius of the cylinder initially centred at $\left(0, y_c\right)$ with angular velocity of $\Omega$. As in this study we consider only fixed walls, the frame moving with the wall is used as the reference frame. By taking $x-\uw t$ instead of $x$ as the axial coordinate and $u-\uw$ instead of $u$ in \eqref{eq:flow}, we obtain the flow for a cylinder moving at $\uc=\uo=-\uw$. If, in addition, we want to impose an oscillation to the cylinder, then we use the superposition principle since \eqref{eq:flow} is solution of the Stokes (linear) equation. To obtain this, we add to \eqref{eq:flow} the velocity field of a uniform flow oscillating in time with velocity $-\beta \cos(\omega t)$. In the frame moving with the wall, the velocity field remains unchanged, and only a change of coordinate is necessary, i.e. $x-\uw t +\beta \sin(\omega t)/\omega$ is used instead of $x$, which results in a flow developing close to a fixed wall around a moving cylinder of velocity $\uc = \uo + \beta \cos(\omega t)$.

\section*{Appendix B}\label{appB}
An infinitesimal perturbation $\boldsymbol{\xi}$ to the trajectory $\xb(t, \xb_0) = (x(t, \xb_0), y(t, \xb_0))$ is described by the following linearised flow
\begin{equation}\label{lin}
\dot{\boldsymbol{\xi}} = \bnabla \ub (\xb(t, \xb_0),t) \boldsymbol{\xi},
\end{equation}
where $\ub=(u,v)$ is the velocity vector defined at time $t$ and at the initial position $\xin$, that will be omitted to simplify the notation. Following \cite{Haller2003}, \cite{Lekien2008} introduced a local coordinate system aligned to the local velocity vector to transform \eqref{lin} into a system that can be solved explicitly. Here, we consider the more general case where the coordinate system has arbitrary axes defined by the tangent unit vector $\eb(\xb(t), t)$ and the normal unit vector $\nb(\xb(t), t)$. With the transformation matrix $\boldsymbol{T}(\xb(t), t)$ defined by
\begin{equation}
\boldsymbol{T} = [\eb(\xb, t) \ \ \nb(\xb, t)],
\end{equation}
the coordinates $\boldsymbol{\xi}$ along $\xb(t)$ are changed to the coordinates $\etab =(\eta_1, \eta_2)$ aligned with $\eb$ and $\nb$:
\begin{equation}
\boldsymbol{\xi} = \boldsymbol{T} (\xb,t) \etab.
\end{equation}
and the linearised flow \eqref{lin} is transformed into
\begin{equation}\label{sys}
\setlength{\arraycolsep}{0pt}
\renewcommand{\arraystretch}{1.3}
\etabp = \left(
\begin{array}{cc}
  \Sp (t) & \quad \gamma(t)  \\[2pt]
  \displaystyle
  \alpha (t) & \quad \Sn (t)
\end{array} \right)\etab.
\end{equation}
In the new system \eqref{sys},
\begin{equation}
\Sp =  \prods{\eb}{\gradu}{\eb}_\xo =  \prodss{\eb}{\boldsymbol{\mathcal{S}}}{\eb}_\xo
\end{equation}
is the tangential rate of strain along the trajectory $\xb (t)$, $\boldsymbol{\mathcal{S}}$ being the deformation tensor. The normal rate of strain is
\begin{equation}
\Sn = \prods{\nb}{\gradu}{\nb}_\xo = -\Sp + \delta(t) \quad \text{with} \quad \delta(t) = \bnabla\bcdot\ub\vert_\xo,
\end{equation}
and the antidiagonal terms are
\begin{equation}
\gamma(t)  = \prods{\eb}{\gradu}{\nb}_\xo + \nb\bcdot\ebp |_\xo
\end{equation}
and
\begin{equation}
\alpha(t) = \prods{\nb}{\gradu}{\eb}_\xo - \nb\bcdot\ebp |_\xo.
\end{equation}

If the local coordinate system is aligned with the local velocity vector, $\eb$ is defined as $\ub/|\ub|$, and $\alpha(t)$ becomes
\begin{equation}
\alpha(t) = -\left. \frac{\ub^\perp \bcdot \ub_t}{|\ub|^2} \right|_\xo, 
\end{equation}
where $\ub^\perp$ is a vector orthogonal to $\ub$. In \cite{Haller2003}, steady and slowly varying velocity fields are studied, so that $\alpha (t) = 0$ or is negligible, and the system \eqref{sys} becomes upper diagonal and can therefore be integrated. In \cite{Lekien2008}, this system is used to describe the separation on a slip boundary. Since the boundary is fixed, $\ub_t$ is parallel to $\ub$, and $\alpha$ vanishes again. Therefore, in both cases, the line $\{\eta_2=0\}$ is an invariant subspace for \eqref{sys}, which means that a vector initially tangent to $\eb$ remains tangent for all $t$. The stability of the line $\{\eta_2=0\}$ can be obtained from the evolution of $\eta_2$ in time. By defining
\begin{equation}\label{lambda}
\lambda(t_0,t) = \int_{t_0}^t \Sp(\tau)\mbox{d}\tau,
\end{equation}
we obtain, for an incompressible flow,
\begin{equation}
\eta_2(t) =\eta_2(t_0)\,\text{e}^{-\lambda(t_0,t) }
\end{equation}
which indicates that for $\lambda > 0$ ($\lambda < 0$), the line $\{\eta_2=0\}$ attracts (repels) over the time interval $[t_0,t]$.

\bibliographystyle{unsrt} 

\end{document}